\documentclass[prd,reprint,nofootinbib,notitlepage,aps,tightenlines,amsmath,amssymb,showpacs,superscriptaddress]{revtex4-1}

\usepackage[hyperindex,breaklinks]{hyperref}
\usepackage{graphicx,epstopdf,amsmath,amsfonts,amssymb,color}

\def\lsim{\mathrel{\rlap{\lower4pt\hbox{\hskip1pt$\sim$}}
    \raise1pt\hbox{$<$}}}                
\def\gsim{\mathrel{\rlap{\lower4pt\hbox{\hskip1pt$\sim$}}
    \raise1pt\hbox{$>$}}}                
\def\be{\begin{equation}}
\def\ee{\end{equation}}
\def\ba{\begin{eqnarray}}
\def\ea{\end{eqnarray}}
\def\ge{\mathrel{\raise.3ex\hbox{$>$\kern-.75em\lower1ex\hbox{$\sim$}}}}
\def\la{\mathrel{\raise.3ex\hbox{$<$\kern-.75em\lower1ex\hbox{$\sim$}}}}

\def\blfootnote{\xdef\@thefnmark{}\@footnotetext}

\def\simgt{\mathrel{\raise.3ex\hbox{$>$\kern-.75em\lower1ex\hbox{$\sim$}}}}
\def\simlt{\mathrel{\raise.3ex\hbox{$<$\kern-.75em\lower1ex\hbox{$\sim$}}}}

\newcommand{\bi}[1]{\bibitem{#1}}

\newcommand{\nc}{\newcommand}

\nc{\gone}{\bar g_{\pi NN}^{(1)}}
\nc{\gzero}{\bar g_{\pi NN}^{(0)}}
\nc{\al}{\alpha}
\nc{\ga}{\gamma}
\nc{\de}{\delta}
\nc{\ep}{\epsilon}
\nc{\ze}{\zeta}
\nc{\et}{\eta}
\nc{\ka}{\kappa}
\nc{\rh}{\rho}
\nc{\si}{\sigma}
\nc{\ta}{\tau}
\nc{\up}{\upsilon}
\nc{\ph}{\phi}
\nc{\ch}{\chi}
\nc{\ps}{\psi}
\nc{\om}{\omega}
\nc{\Ga}{\Gamma}
\nc{\De}{\Delta}
\nc{\La}{\Lambda}
\nc{\Si}{\Sigma}
\nc{\Up}{\Upsilon}
\nc{\Ph}{\Phi}
\nc{\Ps}{\Psi}
\nc{\Om}{\Omega}
\nc{\ptl}{\partial}
\nc{\del}{\nabla}
\nc{\ov}{\overline}
\nc{\newcaption}[1]{\centerline{\parbox{15cm}{\caption{#1}}}}
\nc{\us}{U(1)$_S$}

\def\beq{\begin{equation}}
\def\eeq{\end{equation}}
\def\bmat{\begin{displaymath}}
\def\emat{\end{displaymath}}
\def\bear{\begin{eqnarray}}
\def\eear{\end{eqnarray}}
\def\ba{\begin{eqnarray}}
\def\ea{\end{eqnarray}}
\def\bery{\begin{array}}
\def\ery{\end{array}}
\def\bit{\begin{itemize}}
\def\eit{\end{itemize}}
\def\ben{\begin{enumerate}}
\def\een{\end{enumerate}}
\def\btab{\begin{tabular}}
\def\etab{\end{tabular}}
\def\btbl{\begin{table}}
\def\etbl{\end{table}}
\def\bfig{\begin{figure}[htb]}
\def\efig{\end{figure}}
\def\bpic{\begin{picture}}
\def\epic{\end{picture}}

\def\ga{\mathrel{\raise.3ex\hbox{$>$\kern-.75em\lower1ex\hbox{$\sim$}}}}
\def\la{\mathrel{\raise.3ex\hbox{$<$\kern-.75em\lower1ex\hbox{$\sim$}}}}
\def\gappeq{\mathrel{\rlap {\raise.5ex\hbox{$>$}}
{\lower.5ex\hbox{$\sim$}}}}
\def\lappeq{\mathrel{\rlap{\raise.5ex\hbox{$<$}}
{\lower.5ex\hbox{$\sim$}}}}

\def\gyr{{\rm \, G\kern-0.125em yr}}
\def\mev{{\rm \, Me\kern-0.125em V}}
\def\gev{{\rm \, Ge\kern-0.125em V}}
\def\tev{{\rm \, Te\kern-0.125em V}}

\renewcommand{\bar}{\overline}

\newcommand{\rst}[1]{\raise+.6ex\hbox{#1}}

\begin{document}

\title{EDM Signatures of PeV-scale Superpartners}

\author{David McKeen}
\affiliation{Department of Physics and Astronomy, University of Victoria, 
Victoria, BC V8P 5C2, Canada}

\author{Maxim Pospelov}
\affiliation{Department of Physics and Astronomy, University of Victoria, 
Victoria, BC V8P 5C2, Canada}
\affiliation{Perimeter Institute for Theoretical Physics, Waterloo, ON N2J 2W9, 
Canada}

\author{Adam Ritz}
\affiliation{Department of Physics and Astronomy, University of Victoria, 
Victoria, BC V8P 5C2, Canada}

\date{March 2013}

\begin{abstract}
A possible supersymmetric interpretation of  the new Higgs-like 126~GeV resonance involves a high sfermion mass scale, from tens of TeV to a PeV or above. This scale provides sufficiently large loop corrections to the Higgs mass and can naturally resolve the constraints from flavor-violating observables, even with a generic flavor structure in the sfermion sector. We point out that such high scales could still generate CP-violating electric dipole moments (EDMs) at interesting levels due to the enhancement of left-right (LR) sfermion mixing. We illustrate this by saturating the light fermion mass corrections from the sfermion threshold, leaving the gaugino masses unconstrained. In this framework, we find that the current EDM bounds probe energy scales of  0.1~PeV or higher; this is competitive with the reach of $\epsilon_K$ and more sensitive than other hadronic and leptonic flavor observables. We also consider the sensitivity to higher dimensional supersymmetric operators in this scenario, including those that lead to proton decay.
\end{abstract}
\maketitle

\section{Introduction}
\label{sec:intro}

The recent LHC discovery \cite{higgs} of a 126~GeV resonance with properties consistent with those of the Standard Model (SM) Higgs boson, combined
with the lack of evidence for new weak-scale physics, has cast further doubt on supersymmetry (SUSY) as a natural solution to the hierarchy problem.
While it is tantalizing that the mass of the Higgs-like boson is low enough to be compatible with minimal supersymmetry, it is sufficiently far above the 
tree-level bound to require large loop corrections that point to very heavy sfermions, beyond the reach of the LHC. Of course, one can still contemplate
model scenarios that avoid tuning in the Higgs sector by invoking more complex SUSY spectra that hide some of the sfermions around the weak scale.
However, the indirect hint from the scale of Higgs mass is clearly consistent  with the lack of direct evidence for new physics and the already 
strong indirect constraints from flavor- and CP-violating observables.

In this paper, we will make the assumption that supersymmetry is valid at high scales,  and study what
seems the simplest viable scenario with a tuned Higgs sector and heavy super-partners~\cite{split}.
We will then reconsider the sensitivity of indirect 
probes in this light, specifically the role of searches for CP-violating electric dipole moments (EDMs) and flavor-violating decays. An underlying assumption
will be that technical naturalness remains a valid criterion in the fermion sector (if not for the Higgs itself). Working with high-scale SUSY breaking allows for
a generic flavor structure in the sfermion sector. We will denote the generic threshold scale as $\La_{\rm SUSY}$, corresponding to 
the scale of sfermion and higgsino masses ($M_{\rm sf}\sim\mu\sim\Lambda_{\rm SUSY}$), while we allow for the gauginos with 
mass $M_i$, $i=1,\,2,\,3$, to lie in the range 1~TeV$\leq M_i \leq \La_{\rm SUSY}$.
This takes into account the fact that RG running from high scales may lead to some splitting, or that the mechanism of SUSY breaking can lead to a loop-factor
suppression of $M_i$.

The presence of a general sfermion flavor structure in this framework implies, perhaps counter-intuitively, an enhanced relative sensitivity of certain
flavor-diagonal observables. In particular, electric dipole moments of light fermions require a chirality flip and can be enhanced in the presence of 
${\cal O}(1)$ flavor mixing; for example the up quark EDM can be proportional to $m_t$ in place of $m_u$ \cite{flavored,pr,hnp,abgps,flavored_higgs}. This tends to enhance the importance of 
EDMs as compared to chirality-flipping flavor observables, that usually involve down-type fermions and are not enhanced by $m_t$, or 
chirality-preserving flavor observables. 

The remainder of this paper will be devoted to justifying the above statement in more detail. As noted above, we will insist
on technical naturalness in the radiative corrections to the fermion masses, 
\be
 \de m_u \propto \theta_u^2 m_t \frac{M_3}{\La_{\rm SUSY}} \lsim m_u,
\ee 
where $\theta_u^2$ denotes a combination of flavor mixing angles to be discussed below. Under this constraint, and allowing
for a hierarchy between the gaugino masses $M_i$ and $\La_{\rm SUSY}$, 
we find that fermion EDMs and quark chromo-EDMs (CEDMs) scale as
\begin{align}
 d_f &\sim c_1\frac{\de m_f}{\La_{\rm SUSY}^2} \theta_{\rm CP}, \\
 \tilde{d}_q &\sim c_2\frac{\de m_q}{\La_{\rm SUSY}^2} \ln\left(\frac{M_3^2}{\La_{\rm SUSY}^2}\right)
 \theta_{\rm CP},
\end{align}
with $c_i$ an {\cal O}(1) numerical factor that depends on $M_i/\La_{\rm SUSY}$, and $\theta_{\rm CP}$ the corresponding phase. 
In the absence of any additional constraints on these phases, it follows that current experiments are sensitive to sfermion mass scales in the 0.1~PeV range. 

We will consider two examples which characterize this scenario:

\begin{enumerate}
\item {\it Maximal mixing:} We take the gauginos to be light (TeV-scale), assume large sfermion mixing, and adjust the SUSY scale to saturate 
$\de m_u \propto m_t M_3/\La_{\rm SUSY} \sim m_u$.
\item {\it Maximal threshold:}  We take all superpartners with masses of order $\La_{\rm SUSY}$, and adjust the mixing angles to saturate 
$\de m_u \propto \theta_u^2 m_t \sim m_u$.
\end{enumerate}

In the next section, we discuss the EDM sensitivity in more detail. In Sec.~\ref{sec:flavor}, we contrast the EDM reach with conventional
probes of flavor-violation, with $\ep_K$ being the most sensitive. In Sec.~\ref{sec:highdim}, we also point out that the usual flavor-constraints on dimension-5 sources, such as those leading to proton decay, are relaxed in this scenario as compared to weak-scale SUSY. We finish with some concluding remarks in Sec.~\ref{sec:conclusions}.

\section{Fermion masses and EDMs}
\label{sec:mass_edms}
\begin{figure}[t]
\begin{center}
\includegraphics[width=.38\textwidth]{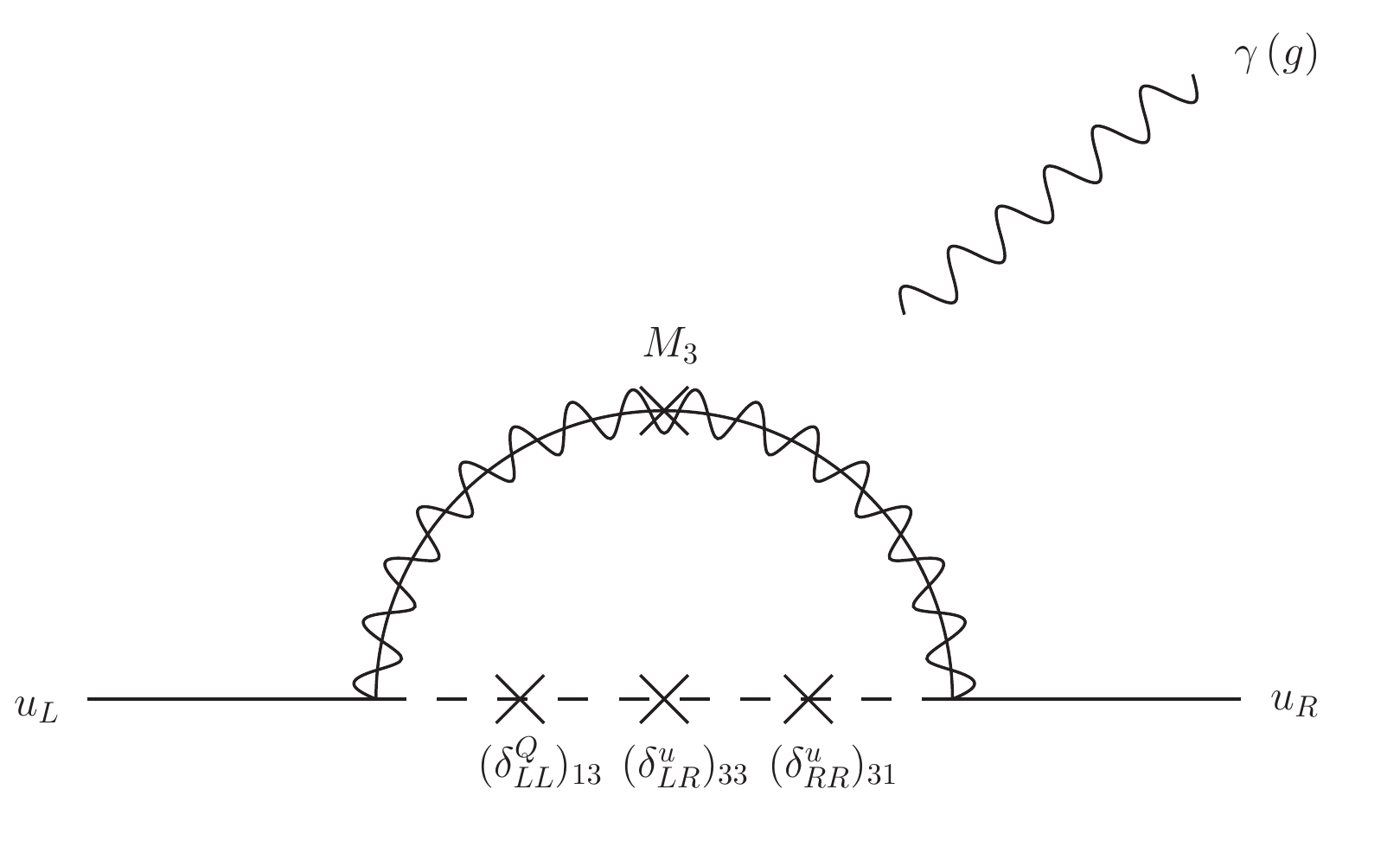}
\caption{The diagram that generates a contribution to the $u$ quark mass, $\delta m_u$, in Eq.~(\ref{eq:umass}). Analogous diagrams can be drawn for the $d$ quark and the electron. Additionally, (C)EDMs are generated by this diagram when a photon (gluon) is attached.}
\label{fig:uloop}
\end{center}
\end{figure}

In the scenario described above, the large top mass can potentially seed the mass of the up quark. In the superCKM basis, with diagonal up and down quark Yukawas
and gaugino mass matrices, it is convenient to use the language of mass insertions (MIs)~\cite{hkr}. We treat MIs as small perturbations, although they can potentially be ${\cal O}(1)$; this is valid at the level of our naturalness-based estimates.
The contribution of a gluino-squark loop, as in Fig.~\ref{fig:uloop}, to the $u$ quark mass is then given by
\begin{align}
\delta m_u&\sim\frac{\alpha_s}{3\pi}f_m(r_3)M_3 \big(\delta^Q_{LL}\big)_{13}\big(\delta^u_{LR}\big)_{33}\big(\delta^u_{RR}\big)_{31}.
\label{eq:umass}
\end{align}
In this formula, $r_i\equiv M_i^2/\La_{\rm SUSY}^2$, where $\La_{\rm SUSY}$ is the common diagonal LL and RR squark mass scale, 
and $f_m(r)$ is a loop function with the following limits in the two cases discussed in Section~1,
\be
 f_m(r) \rightarrow \left\{ \begin{array}{cc} \raisebox{0.5ex}{$\frac{2}{3},$} & \raisebox{0.5ex}{$r\ll 1 \; ({\rm case\,1}),$} \\ \raisebox{-0.5ex}{$\frac{1}{6},$} & \raisebox{-0.5ex}{$r=1\; ({\rm case\,2}).$} \end{array} \right.
\ee
The off-diagonal LL and RR mass insertions are defined as the corresponding entry in the $M_{LL}^2$ and $M_{RR}^2$  mass matrices, weighted by
$\La_{\rm SUSY}^{-2}$. Finally, the LR insertion is $\La_{\rm SUSY}^{-2}\mu m_t \cot\beta $, where we consider the case of small $A$ terms, $A\ll \La_{\rm SUSY}$.

With squarks at the 100--1000~TeV scale, the mixing can potentially be large, $\theta_{u13}^2\equiv(\delta^Q_{LL})_{13}(\delta^u_{RR})_{31}\sim{\cal O}\left(1\right)$. A universality assumption at high scales would not generally forbid large LL mixing to arise through RG evolution, but large RR mixing would
require a more generic flavor structure even at high scales.\footnote{See \cite{flavored,hnp} for analyses with LL or RR mixing in specific models.} To account for both cases 1 and 2 discussed in the previous section, we will present
the results below in terms of the combination $\theta_f^2 M_i$, which takes the benchmark value of 300~GeV for both examples. For case 1, we consider
$M_i\sim 1$~TeV with $\theta_f^2\sim 1/3$, while for case 2 we have $M_i \sim \La_{\rm SUSY}$ with $\theta_f^2 \ll 1$. 

In the case of corrections to the up quark mass, we obtain
\begin{align}
\delta m_u&\sim\frac{\alpha_s}{3\pi}f_m(r_3) \theta_{u13}^2\frac{m_t M_3}{\Lambda_{\rm SUSY}\tan\beta} 
\\
&\simeq 1.5~{\rm MeV}f_m(r_3){\left(\frac{4}{\tan\beta}\right)}{\left(\frac{\theta_{u13}^2 M_3}{300~\rm GeV}\right)}{\left(\frac{100~\rm TeV}{\Lambda_{\rm SUSY}}\right)}, \nonumber
\end{align}
where as above $\La_{\rm SUSY}$ denotes the common squark and higgsino mass, and we have retained just the term proportional to $\mu\sim \La_{\rm SUSY}$ in the squark LR mass insertion (assuming that the trilinear terms are parametrically smaller as noted above, $A\ll \La_{\rm SUSY}$). The scales have been adjusted so that, for moderate $\tan\beta$ (as suggested by a 126~GeV Higgs with a high SUSY scale), this contribution is of the right order of magnitude to saturate the $u$ quark mass, $\delta m_u\sim m_u$, normalized at this high scale. Equivalently, for the hierarchical spectrum  in case 1, no tuning of the mixing angles $\theta_{u13}^2$ is required to keep the $u$ quark light.

In the case of the $d$ quark, the mass shift is given by a similar expression,
\begin{align}
\delta m_d&\sim\frac{\alpha_s}{3\pi}f_m(r_3)\theta_{d13}^2\frac{m_b M_3\tan\beta}{\Lambda_{\rm SUSY}}
\\
&\simeq 0.5~{\rm MeV}f_m(r_3){\left({\frac{\tan\beta}{4}}\right)}{\left(\frac{\theta_{d13}^2 M_3}{300~\rm GeV}\right)}{\left(\frac{100\,\rm TeV}{\Lambda_{\rm SUSY}}\right)},
\nonumber
\end{align}
where we have defined $\theta_{d13}^2\equiv(\delta^Q_{LL})_{13}(\delta^d_{RR})_{31}$. For the chosen parameters, this is likely too small a contribution to saturate the full $d$ quark mass. However, as in the $u$ quark case, for the hierarchical spectrum in case 1 there is no need to tune the mixing angles $\theta_{d13}^2$ in order to keep 
the $d$ quark light. See, e.g.,~\cite{masses_old} for scenarios where  some (or all) of the down-type fermion masses and quark mixing angles arise 
from SUSY threshold corrections.

For the electron, a similar mass correction arises at the SUSY threshold in this scenario. The  leading contribution is given by
\begin{align}
\delta m_e&\sim\frac{\alpha_1}{4\pi}f_m(r_1)M_1\big(\delta^L_{LL}\big)_{13}\big(\delta^e_{LR}\big)_{33}\big(\delta^e_{RR}\big)_{31} \nonumber\\
&\sim\frac{\alpha}{4\pi\cos^2\theta_W}f_m(r_1)\theta_{e13}^2\frac{m_\tau M_1\tan\beta}{\Lambda_{\rm SUSY}}
\\
&\simeq0.02\,{\rm MeV}f_m(r_1){\left(\frac{\tan\beta}{4}\right)}{\left(\frac{\theta_{e13}^2 M_1}{300\,\rm GeV}\right)}{\left(\frac{100\,\rm TeV}{\Lambda_{\rm SUSY}}\right)},
\nonumber
\end{align}
where $\theta_{e13}^2\equiv(\delta^L_{LL})_{13}(\delta^e_{RR})_{31}$, which requires no tuning given the hierarchical spectrum of case 1. Unless $\tan\beta$ is very large, this contribution in case 1 is always somewhat smaller than the full  electron mass.

We now turn to CP-violating observables, and their sensitivity to the threshold scale. Firstly, note that
imaginary corrections to the quark masses also renormalize the QCD vacuum angle $\bar{\theta}$, 
\begin{align}
 \delta \bar{\theta} &\sim \frac{{\rm Im}(m_u)}{m_u} \sim -\frac{\al_s}{3\pi}f_m(r_3) \theta_{u13}^2  \frac{m_tM_3}{m_u\tan\beta \La_{\rm SUSY}} \sin\theta_{\tilde{u}\mu}, \nonumber
 \\
  &\simeq~ 0.6 f_m(r_3) \times{\left(\frac{4}{\tan\beta}\right)}{\left(\frac{\theta_{u13}^2M_3}{300\,\rm GeV}\right)}{\left(\frac{100\,\rm TeV}{\Lambda_{\rm SUSY}}\right)}
  \nonumber\\
  &\qquad\qquad \times {\left(\frac{\sin\phi_{\tilde u\mu}}{1/\sqrt{2}}\right)},
\end{align}
where $\phi_{\tilde u\mu}$ denotes a linear combination of the basis-invariant phases in the off-diagonal up squark mass matrix elements, and the
relative phase between $\mu$ and the gluino mass. This leads to a correction that is ${\cal O}(10^{10})$ times too large, given the limit on the neutron EDM \cite{nedm}, unless the mixing angles are correspondingly suppressed. We will instead assume
as usual that the vacuum angle is relaxed to zero via the axion mechanism. 

This still leaves a number of higher dimension CP-odd EDM sources,
and we will focus on the chromo-EDMs\footnote{With squarks much heavier than gluinos, the CEDM is logarithmically enhanced relative to the EDM.} which are sensitively probed by the current constraint on the neutron and Hg EDMs \cite{nedm,Griffith:2009zz}.
Given the shift in the $u$ quark mass arising from the gluino-squark loop in Fig.~\ref{fig:uloop}, a contribution to its (C)EDM arises from attaching a photon (gluon) to this diagram. In either case 1 or 2, the $u$ quark CEDM can be written in the form,
\begin{align}
\tilde d_u &\sim f_q(r_3)\frac{\delta m_u}{\Lambda_{\rm SUSY}^2}\sin\phi_{\tilde u\mu},
\end{align}
where $\phi_{\tilde u\mu}$ as above denotes a linear combination of the basis-invariant phases in the loop. The function $f_q(r)$ denotes the ratio of the loop
function that enters the CEDM calculation~\cite{cedm,hnp,abgps} relative to $f_m(r)$, and takes the form,
\be
  f_q(r) \rightarrow \left\{ \begin{array}{cc} \raisebox{0.5ex}{$\frac{27}{8}\ln(r), $}& \raisebox{0.5ex}{$r\ll 1 \; ({\rm case\,1}),$} \\ \raisebox{-0.5ex}{$-\frac{11}{40},$} & \raisebox{-0.5ex}{$r=1\; ({\rm case\,2}).$} \end{array} \right. 
\ee
If we focus on case 1, with a hierarchical spectrum of gaugino and sfermion masses, we find the result
\begin{align}
\tilde{d}_u&\simeq 5{\times}10^{-26}\,{\rm cm}{\left(\frac{4}{\tan\beta}\right)}{\left(\frac{\theta_{u13}^2M_3}{300\,\rm GeV}\right)}{\left(\frac{100\,\rm TeV}{\Lambda_{\rm SUSY}}\right)^3} \nonumber
\\
&\qquad\qquad\times\left[\ln\left(\frac{\Lambda_{\rm SUSY}^2}{M_3^2}\right)\Big/10\right]\left(\frac{\sin\phi_{\tilde u\mu}}{1/\sqrt{2}}\right).
\end{align}
For case 2, the result is smaller: the log enhancement is absent, and the numerical coefficient is also an order of magnitude smaller than in case 1,
$f_q(r=1)/f_q(r\simeq 10^{-6})\sim O(10^{-2})$. 

As in the case of the mass shifts, we can write a similar expression for the $d$ quark CEDM,
\begin{align}
\tilde d_d&\sim f_q(r_3)\frac{\delta m_d}{\Lambda_{\rm SUSY}^2}
\sin\phi_{\tilde d\mu}
\\
&\simeq 2{\times}10^{-26}\,{\rm cm}{\left(\frac{\tan\beta}{4}\right)}{\left(\frac{\theta_{d13}^2M_3}{300~\rm GeV}\right)}{\left(\frac{100~\rm TeV}{\Lambda_{\rm SUSY}}\right)^3} \nonumber
\\
&\qquad\qquad\times\left[\ln\left(\frac{\Lambda_{\rm SUSY}^2}{M_3^2}\right)\Big/10\right]\left(\frac{\sin\phi_{\tilde d\mu}}{1/\sqrt{2}}\right), \nonumber
\end{align}
where in the second line we have again focused on case 1, with $\phi_{\tilde d\mu}$ defined analogously.

The CEDMs of $u$ and $d$ quarks are presently best probed by the limit on the mercury EDM, $\left|d_{\rm Hg}\right|<3.1\times10^{-29}~e\, {\rm cm}$~\cite{Griffith:2009zz}. This translates into a limit on the quark CEDMs, $|\tilde d_u-\tilde d_d|\lesssim 6\times10^{-27}~{\rm cm}$,\footnote{An orthogonal combination of CEDMs is also constrained, with different hadronic and nuclear uncertainties, by the current limit on the EDM of the neutron \cite{nedm}.} 
 implying that in this scenario the  mercury EDM can currently access SUSY scales of
\begin{align}
\Lambda_{\rm SUSY}\sim 200~{\rm TeV}{\left(\frac{\theta_{d13}^2M_3}{300~\rm GeV}\right)}^{1/3}\left(\frac{|\sin\phi_{\tilde q\mu}|}{1/\sqrt{2}}\right)^{1/3},
  \end{align}
  for moderate values of $\tan\beta$.

As with the quarks, the electron receives a contribution to its EDM by attaching a photon to the same diagram that is responsible for the mass shift,
\begin{align}
d_e&\sim e f_e(r_1)\frac{\de m_e}{\La_{\rm SUSY}^2} \sin\phi_{\tilde e\mu}
\\
&\simeq 1{\times}10^{-29}\,e\,{\rm cm}{\left(\frac{\tan\beta}{4}\right)}{\left(\frac{\theta_{e13}^2M_1}{300~\rm GeV}\right)} \nonumber\\
& \qquad\qquad \times {\left(\frac{100~\rm TeV}{\Lambda_{\rm SUSY}}\right)^3}{\left(\frac{\sin\phi_{\tilde e\mu}}{1/\sqrt{2}}\right)}, \nonumber
\end{align}
where the second second line again follows for case 1, and the function $f_e(r)$ takes the form \cite{edm,hnp,abgps},
\be
  f_e(r) \rightarrow \left\{ \begin{array}{cc} \raisebox{0.5ex}{$\frac{3}{4}, $}& \raisebox{0.5ex}{$r\ll 1 \; ({\rm case\,1}),$} \\ \raisebox{-0.5ex}{$\frac{1}{5},$} & \raisebox{-0.5ex}{$r=1\; ({\rm case\,2}).$} \end{array} \right. 
\ee
For the chosen normalization parameters, this is significantly below the current constraint of $\left|d_e\right|\lesssim 1.05\times10^{-27}~e\,{\rm cm}$~\cite{Hudson:2011zz}, unless $\tan\beta$ is particularly large. Notice also that the 1-loop bino-slepton EDM diagram does not receive a logarithmic
enhancement. The technical reason for the log-enhancements of the CEDMs at one loop can be traced to the fact that the gluino carries a color charge,
and more precisely to the part of the gluino propagator given by $ t^a G^a_{\mu\nu} \sigma_{\mu\nu} M_3 /(p^2-M_3^2)^{2}$ in an external field; the corresponding term in the bino propagator is absent due to its neutrality. A similar log-enhancement does appear in the chargino-slepton loop, but given that 
one of the vertices is proportional to the Yukawa coupling of the external fermion, such diagrams are subleading as they do not recieve the $m_\tau/m_e$ enhancement
due to large LR mixing.

\begin{figure}[t]
\begin{center}
\includegraphics[width=.48\textwidth]{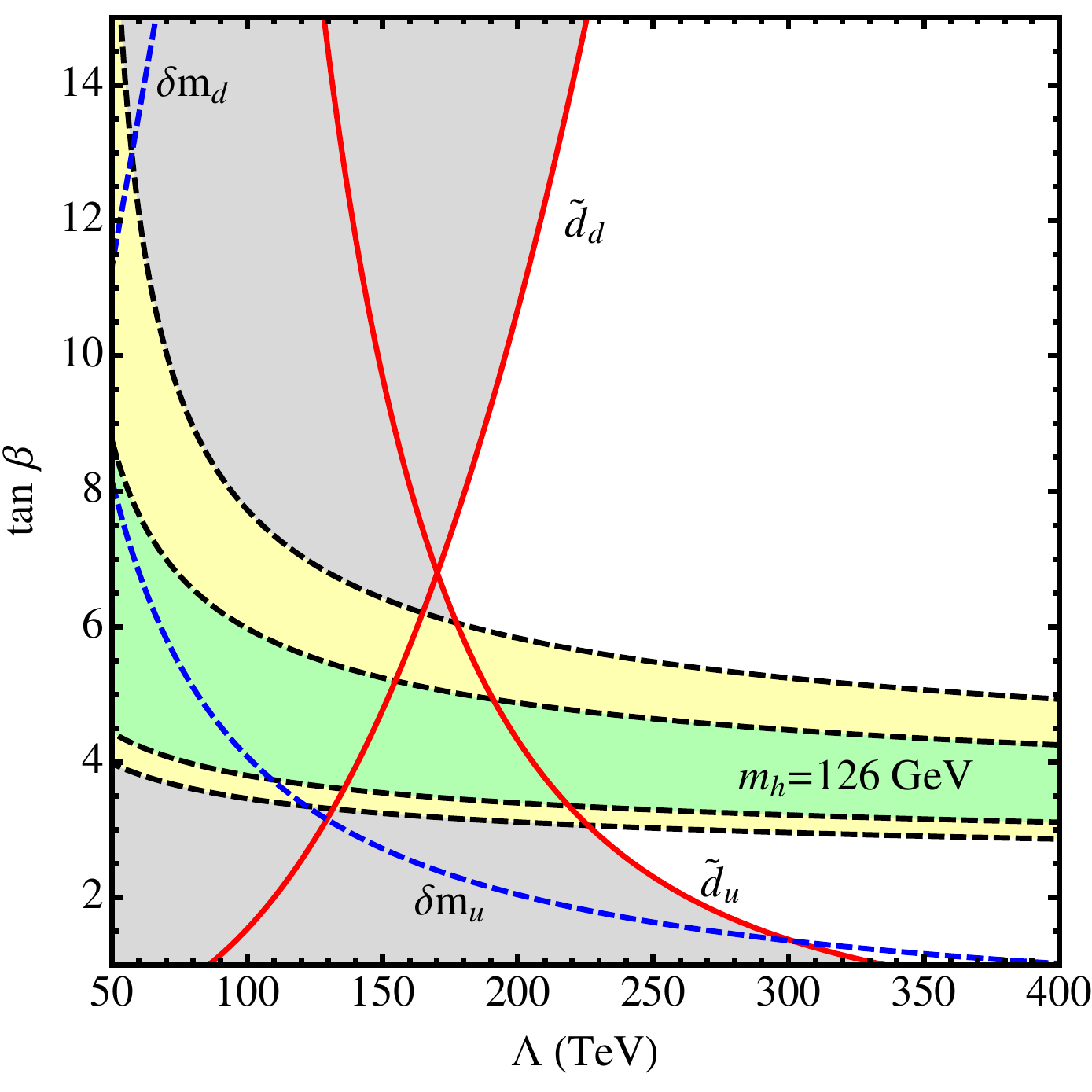}
\caption{Contours of $\delta m_u=1\,{\rm MeV}$ and $\delta m_d=2\,{\rm MeV}$ (blue, dashed) and $\tilde d_q=6\times10^{-27}$~cm for $q=u,\,d$ (red, solid) are shown, with $\theta_{q13}^2=1/3$, $M_{3}=1$~TeV, and $\sin\phi_{\tilde q\mu}=1/\sqrt{2}$. If the limit $|\tilde d_u-\tilde d_d|\lesssim 6\times10^{-27}~{\rm cm}$ from the mercury EDM~\cite{Griffith:2009zz}  is interpreted as a limit on $\tilde d_u(\theta_{\tilde{u}\mu})$ and $\tilde d_d(\theta_{\tilde{d}\mu})$ independently, given the distinct CP phases, then the shaded region to the left of each contour is ruled out. For comparison, we have shown the region of parameter space consistent at 2$\sigma$ with a Higgs mass $m_h=125.7\pm0.8$~GeV~\cite{higgs} and the top mass fixed to $m_t=173.5$~GeV (green, inner band) and with $m_t$ in the range $173.5\pm1$~GeV (yellow, outer band). (The one-loop leading-log corrections~\cite{1loop} to the Higgs mass are used here; two-loop corrections tend to lower the band to slightly smaller values of $\tan\beta$, see, e.g.,~\cite{split}.)}
\label{fig:edm}
\end{center}
\end{figure}

If we fix the mixings in the $u$-, $d$-, and $e$-sectors to $\theta_{u,d,e}^2=1/3$ as well as the gaugino masses to $M_{1,3}=1$~TeV, we can calculate the (C)EDMs $\tilde d_{u,d}$, $d_e$ as functions of $\tan\beta$ and $\Lambda_{\rm SUSY}$. In Fig.~\ref{fig:edm}, we show contours of constant $\delta m_q$ and $\tilde{d}_{u,d}$, varying $\tan\beta$ and $\Lambda_{\rm SUSY}$. We see that the EDM limits probe scales of ${\cal O}(0.1)$~PeV or even higher in this scenario. The corresponding contour for $d_e$ is similar in shape to that for $\tilde{d}_d$, and using the current limit from 
the bound on the EDM of YbF \cite{Hudson:2011zz}, is sensitive to scales of ${\cal O}(30)$~TeV with the same parameters.

Finally, we will comment briefly on the contribution of two-loop Barr-Zee-type diagrams with a closed chargino loop attached to the quark lines by 
a virtual $h\gamma$ pair \cite{BZ}. For the scenarios we are considering, these diagrams constitute a small correction (although they can be important in scenarios where the $h\to\gamma\gamma$ rate is increased through CP-violating interactions~\cite{McKeen:2012av}). In fact, while these corrections
are small for case 1, they are tiny for case 2. Although they have a milder decoupling with $\Lambda_{\rm SUSY}$, 
$d_i^{\rm 2-loop~BZ}\propto 1/\mu \sim 1/\Lambda_{\rm SUSY}$,  they do not receive the corresponding mass enhancement by $m_t/m_u$, 
{\em i.e.} $d_i^{\rm 2-loop~BZ} \sim m_i$, which renders them subdominant over the full range of $\Lambda_{\rm SUSY}$ that is of interest here. 

\section{Flavor-violating observables}
\label{sec:flavor}

In contrast to EDMs, most flavor-violating observables arise in the down-type fermion sector and so cannot access the large $m_t/m_u$ enhancement
from generic flavor mixing at the sfermion scale. Nonetheless, dipole transitions can still be important, particularly for large $\tan\beta$. Observables which
do not require a chirality flip are again comparatively weaker in this scenario, but we still find that $\ep_K$ provides the best sensitivity in the 1-2--sector, 
albeit only probing slightly higher scales than EDMs.

\subsection{Kaon mixing and \boldmath{$\epsilon_K$}}
\label{sec:epsK}
As always, limits from $K^0-\bar K^0$ mixing are extremely important, in particular the constraint from indirect CP-violation in neutral kaon decay. For  case 1, $\ep_K$ takes the form~\cite{flavor}
\begin{align}
\epsilon_K^{\rm SUSY}&=\frac{{\rm Im}\langle \bar K^0\left|{\cal H}_{\rm SUSY}\right| K^0\rangle}{\sqrt{2}\Delta M_K} \nonumber
\\
&\simeq -0.15\left(\frac{100~{\rm TeV}}{\Lambda_{\rm SUSY}}\right)^2{\rm Im}\left\{{\left[\big(\delta^Q_{LL}\big)_{12}^2{+}\big(\delta^d_{RR}\big)_{12}^2\right]}\right.
\nonumber\\
&\;\left.{+}\frac{2}{11}{\left[3{-}2\left(\frac{M_K}{m_s{+}m_d}\right)^2\right]}{\big(\delta^Q_{LL}\big)_{12}}{\big(\delta^d_{RR}\big)_{12}}\right\},
\end{align}
assuming that the $K^0-\bar K^0$ mass difference is dominantly accounted for by the SM. For case 2, the coefficient 0.15 is replaced by 0.30 in the above expression. If all of the squark mass mixings have comparable magnitudes and phases,
\begin{align}
\big(\delta^Q_{LL}\big)_{12}^2\sim\big(\delta^d_{RR}\big)_{12}^2\sim\big(\delta^Q_{LL}\big)_{12}\big(\delta^d_{RR}\big)_{12}\sim \theta_{d12}^2e^{i\phi_{d12}},
\end{align}
then this becomes
\begin{align}
\epsilon_K^{\rm SUSY}&\simeq 0.09\left(\frac{100~{\rm TeV}}{\Lambda_{\rm SUSY}}\right)^2\left(\frac{\theta_{d12}^2}{1/3}\right)\left(\frac{\sin\phi_{\tilde d}}{1/\sqrt 2}\right),
\end{align}
which is relatively insensitive to the gluino mass due to the kinematics of the box diagram.
Requiring that $\ep_K^{\rm SUSY}$ is less than $2.3\times10^{-3}$, limits the SUSY scale as follows,
\begin{align}
\Lambda_{\rm SUSY}\gtrsim 600~{\rm TeV}\left(\frac{\theta_{d12}^2}{1/3}\right)^{1/2}\left(\frac{|\sin\phi_{\tilde d}|}{1/\sqrt 2}\right)^{1/2}.
\end{align}
For case 2, this bound is slightly stronger by a factor of $\sim\sqrt 2$. The stability of this limit under variations in the gaugino masses, contrasts with the enhanced sensitivities of the EDMs in the hierarchical regime. Indeed, for a spectrum of the form given by case 1, the EDM sensitivity approaches that of $\ep_K$. 

For completeness, we note that the bounds from other quark flavor-violating observables, such as $\Delta M_K$, $\ep^\prime/\ep$, $D$, and $B_{d,s}$ mixing, are all weaker in this scenario, setting a bound on the the SUSY scale in the tens of TeV range.

\subsection{Lepton flavor violation}
\label{sec:mu2e}
In the present scenario, with large flavor mixing at the sfermion mass scale, the sensitivity of lepton flavor violating (LFV) decays is somewhat weaker. We begin by discussing $\mu\to e$ conversion in titanium. This can proceed through a box diagram that generates the chirality conserving transition
$\mu q\to eq$ with a branching ratio \cite{flavor,abgps}
\begin{align}
{\cal B}\left(\mu\to e\right)_{\rm Ti,\,box}&\sim10^{-16}\left(\frac{100~{\rm TeV}}{\Lambda_{\rm SUSY}}\right)^4\left(\frac{\theta_{e12}^2}{1/3}\right),
\end{align}
where $\theta_{e12}$ represents the typical LL or RR slepton mixing in the 1-2--sector.

In addition to the chiraility-preserving box diagrams there is also the possibility that a chirality-flipping transition dipole is generated, leading to the LFV decay $\mu\to e\gamma$ as well as $\mu\to e$ conversion. In case 1, where large mixings are conceivable, the amplitude for this transition through bino-slepton exchange can be enhanced by a factor of $m_\tau/m_\mu$, which gives a branching for $\mu\to e\gamma$ of the form \cite{flavor,abgps},
\begin{align}
{\cal B}\left(\mu\to e\gamma\right)&\sim\frac{3\pi\alpha^3\tan^2\beta}{2\cos^4\theta_W}\frac{m_\tau^2}{m_\mu^2}\theta_{e12}^4\frac{M_1^2}{G_F^2\Lambda_{\rm SUSY}^6}
\\
&\simeq 1{\times}10^{-17}{\left(\frac{\tan\beta}{4}\right)^2}{\left(\frac{100\,{\rm TeV}}{\Lambda_{\rm SUSY}}\right)^6} \nonumber
\\
&\qquad\qquad\times{\left(\frac{\theta_{e12}^2M_1}{300\,\rm GeV}\right)^2}, \nonumber
\end{align}
with $\theta_{e12}$ denoting a combination of LL and RR slepton mixing angles, $(\delta^e_{RR})_{23}(\delta^L_{LL})_{31}\sim(\delta^L_{LL})_{23}(\delta^e_{RR})_{31}\sim\theta_{e12}^2$. As for $\mu\to e$ conversion, this transition dipole would give rise to a suppressed branching in Ti at roughly the level
\begin{align}
{\cal B}\left(\mu\to e\right)_{\rm Ti,\,dip.}&\sim4\times10^{-20}{\left(\frac{\tan\beta}{4}\right)^2}{\left(\frac{100\,{\rm TeV}}{\Lambda_{\rm SUSY}}\right)^6} \nonumber
\\
&\qquad\qquad\times{\left(\frac{\theta_{e12}^2M_1}{300\,\rm GeV}\right)^2}.
\end{align}

These LFV rates are significantly below the current limits at the $10^{-12}$ level on ${\cal B}\left(\mu\to e\gamma\right)$~\cite{Adam:2011ch} and on ${\cal B}\left(\mu\to e\right)$ in Ti~\cite{Dohmen:1993mp}. The Mu2e collaboration hopes to improve the $\mu\to e$ reach by four orders of magnitude~\cite{Dukes:2011zz} which could bring it into interesting territory in this scenario.

\section{Higher dimensional SUSY thresholds}
\label{sec:highdim}
\subsection{Proton decay}
\label{sec:protondec}
\begin{figure}[t]
\begin{center}
\includegraphics[width=.35\textwidth]{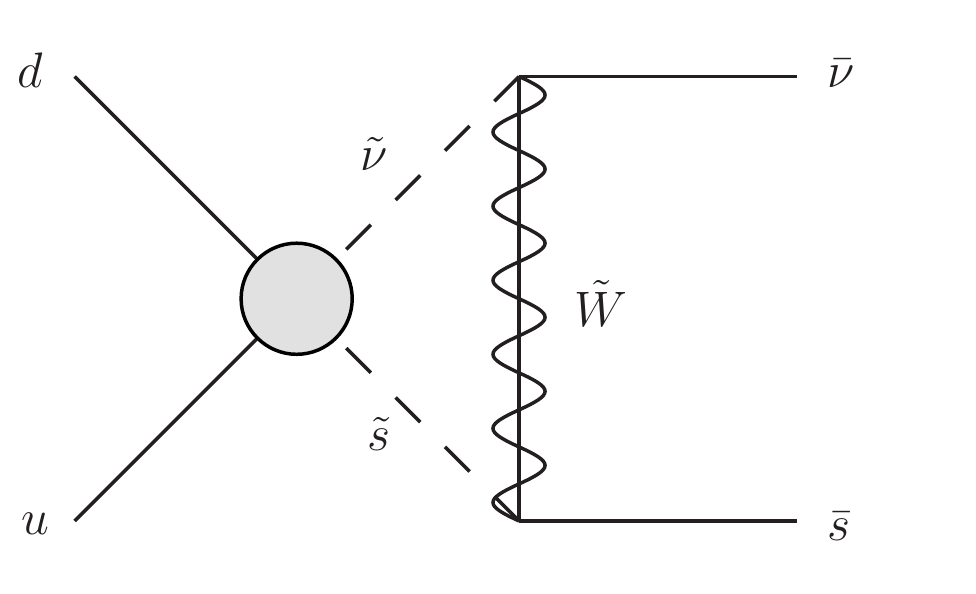}
\caption{A diagram that leads to $p\to K^+\bar\nu$. The shaded blob represents the dimension-five operator that results from the term $QQQL$ in the superpotential in Eq.~(\ref{eq:dim5}), which is dressed by a wino exchange to generate the effective dimension-six proton decay operator.}
\label{fig:protondec}
\end{center}
\end{figure}

Having gaugino masses suppressed relative to those of sfermions can also have an impact on nucleon lifetimes. Proton decay can be problematic even in $R$-parity conserving SUSY models because of dimension-five operators that come from the following terms in the superpotential~\cite{pdecay},
\begin{align}
W\supset \frac{1}{\Lambda_{5L}}QQQL+\frac{1}{\Lambda_{5R}}UUDE.
\label{eq:dim5}
\end{align}
These terms give rise to interactions of the form $qq\tilde q\tilde \ell$, which, when combined with gaugino or Higgsino exchange, lead to the decay of a nucleon, as seen in Fig~\ref{fig:protondec}. In typical SUSY grand unified theories (GUTs) the operators in (\ref{eq:dim5}) are generated by the exchange of color-triplet Higgses. The choice of representation for the Higgses in the theory dictates the structure of these operators and normally the dominant channel for proton decay is $p\to K^+\bar\nu$.

The limit on the proton lifetime in the $K^+\bar\nu$ channel of $3.3\times10^{33}$~years~\cite{superk} leads to impressive bounds on the scales of the operators in (\ref{eq:dim5}),
\begin{align}
\Lambda_{5L}&\gtrsim 10^{22}~{\rm GeV}\left(\frac{100~{\rm TeV}}{\Lambda_{\rm SUSY}}\right)^2\left(\frac{M_\lambda}{\rm TeV}\right),
\label{eq:lambda5L}
\\
\Lambda_{5R}&\gtrsim 10^{21}~{\rm GeV}\left(\frac{100~{\rm TeV}}{\Lambda_{\rm SUSY}}\right)^2 \nonumber\\
& \qquad\times \left(\frac{\mu}{100~\rm TeV}\right)\left(\frac{V_{ts}y_ty_\tau}{4\times10^{-4}}\right).
\end{align}
Note that $\mu$ and the Yukawas appear in the limit on $\Lambda_{5R}$ because obtaining  $K^+\bar\nu$ in the final state from the $UUDE$ operator requires a Higgsino exchange.

The strength of these limits causes considerable tension (often considered as part of the doublet-triplet splitting problem), ruling out the minimal SU(5) SUSY GUT for example~\cite{su5susy}, and posing model building challenges more generally~\cite{susygutmodels}. However, a large splitting between the gaugino masses, which we have labeled here as $M_\lambda$, and the SUSY scale, $\Lambda_{\rm SUSY}$, softens the limit on $\Lambda_{5L}$. 
Compared to weak-scale SUSY, where gauginos and sfermions have a common TeV mass scale, the heavy sfermion case 
introduces an additional suppression factor in the amplitude $\sim{\rm TeV}\times M_\lambda/\Lambda_{\rm SUSY}^2$. This
allows the tension with the limits on proton decay to be eased somewhat and brings the bound on the operator involving gauge couplings closer to that involving the Yukawas.\footnote{If there are large mixings in the sfermion sector, some of the conclusions about the dominant final state can change. In particular, the $UUDE$ operator can lead to $p\to e^+ \pi^0$ through bino exchange. The limit on the scale $\Lambda_{5R}$ in this case would be similar to that on $\Lambda_{5L}$ as shown in (\ref{eq:lambda5L}).}

\subsection{Other dimension-five operators}
\label{sec:dim5}
If one considers the MSSM as an effective theory receiving corrections from multiple thresholds at the level of the superpotential, then additional 
operators of dimension-five have to be taken into account, namely $QULE$, $(H_uH_d)^2$, etc., suppressed by another threshold scale $M$~ \cite{dim5,other_dim5}.
 It was shown in Ref.~\cite{dim5} that EDM constraints on such operators can be particularly strong, limiting the flavor-democratic thresholds to $10^8$ GeV in some cases with the assumption of
weak-scale SUSY. If the SUSY-breaking scale $\Lambda_{\rm SUSY}$ is indeed very large, as considered in this paper, all the corresponding constraints will be 
relaxed by the same relative factor, $\sim {\rm weak~scale}\times M_\lambda/\Lambda_{\rm SUSY}^2$, as discussed above. As a consequence, the constraints on $M$ may not be that 
different from the sfermion mass scale $\Lambda_{\rm SUSY}$.

\section{Concluding remarks}
\label{sec:conclusions}
In this paper, we have argued that conventional indirect probes of new physics can be usefully re-interpreted in light of the discovery of a Higgs-like 126~GeV resonance. In particular, if supersymmetry is realized in nature at all, the need for large radiative contributions to the Higgs mass points to a high SUSY threshold at the 
PeV-scale or above, which is of course consistent with the lack of evidence for new physics. While this may appear to be a disappointing conclusion, it presents a new light
on the threshold itself,  allowing for a generic flavor structure, and perhaps even a theory of flavor. In such a scenario, 
while the importance of flavor-violating observables is well-known, we have emphasized that flavor-diagonal observables actually become comparatively more competitive
due to a significant reduction in chirality suppression. We illustrated this by saturating the mass corrections to light quarks, consistent with naturalness in the fermion mass sector, and then analyzed the ensuing reach of precision measurements. In the presence of a hierarchy between the gaugino masses and the SUSY scale, the CEDMs of
quarks receive an additional logarithmic enhancement. As a consequence, the current EDM limits in the up quark sector exhibit a similar sensitivity to a new SUSY/flavor
threshold as $\ep_K$. It is important to note that EDMs are one of the few precision observables that have significant prospects  for further experimental progress, and 
can be expected to play a more significant role in the future even if future LHC searches do not find new physics sitting at or close to the weak scale.

{\it Note:} As this work was being finalized, Ref.~\cite{moroi} appeared on the arXiv. This work focuses on LFV processes and the electron EDM and reaches similar conclusions.

\acknowledgments{This work was supported in part by NSERC, Canada, and research at the Perimeter Institute is supported in part by the Government of Canada through NSERC and by the Province of Ontario through MEDT. }

\end{document}